# Can the natural system of viruses reconcile the current taxonomy with an alternative classification useful to clinicians?

Alexandr A. Ezhov

State Research Center of Russian Federation
Troitsk Institute for Innovation and Fusion Research,
108840, Troitsk, Moscow, Russia

ezhov@triniti.ru

## Abstract

In 2022, a group of basic and clinical virologists, bioinformaticians, and evolutionary and structural biologists met in Oxford, UK, to develop a consensus on methodologies used to classify viruses. They concluded that virus taxonomy, which is hierarchical and based on evolution, is only one of many possible ways to classify viruses. This taxonomy, while satisfying the four principles they set out, faces difficulties in coordinating with other classification systems useful to clinicians, infectious disease specialists, agronomists, etc. One example discussed is the grouping of different viral strains that cause different diseases into the species *Enterovirus C*. Here we show that the use of a previously proposed variant of a natural virus classification system based on the use of Neural Replicator Analysis can resolve this contradiction by establishing the fine structure of the *Enterovirus C* species, in which strains that cause different diseases are placed in several different cells of the binomial table of viruses. A key element in enabling this is the sophisticated preprocessing of the original viral genomes using neural replicators.

## Introduction

An expert group convened by the International Committee on Taxonomy of Viruses (ICTV) in 2016 debated and affirmed a policy to allow viruses known from their genome sequences alone to be incorporated into virus taxonomy [1]. This policy enables taxonomic assignments without requiring prior knowledge of a virus phenotypic properties, such as host range, type of disease, virus vector, pathogenicity, etc. Nevertheless the use of only non-processed genome sequences and their compairement after alignment leads to the situations when highly clinically diverse viruses are characterized as belonging to the same species because their genomes differ in only a small degree. Actually, it motivates virologists to separate their taxonomic systems from classification suited for clinicysts, plant biologists, etc.

In 2022 a group of basic and clinical virologists, bioinformaticians, and evolutionary and structural biologists met in Oxford, United Kingdom, to develop a consensus on methodologies used for virus classification [2]. They formulated four principles of virus taxonomy: 1) virus taxa should be monophyletic; 2) phenotypic and ecological properties of viruses may inform, but not override, evolutionary relatedness in the placement of ranks; 3) alternative classifications that consider phenotypic attributes, such as being vector-borne, infecting a certain type of host or displaying specific pathogenicity, may serve important clinical and regulatory purposes but often create polyphyletic categories that do not reflect evolutionary relationships; 4) evolution based framework enables viruses discovered by metagenomics to be incorporated into the ICTV taxonomy under suitable control [2].

The second principle suggests that such classification schemes based on the use of phenotypic and ecological features cannot be considered or called taxonomies. It was already mentioned in [3] that in *natural virus system* these properties should be directly derived from the place of the virus genome in the system which can have the form of a table. It was also shown that such properties as virus vectors can be really derived to some extent from their place in the table for the genus *Flavivirus* [4]. The type of disease is defined by the position in the table for the cases of *Human papillomaviruses* and *Caulimoviruses*, and the viral hosts is defined by the position in the table for the *Caulimoviruses* and *Polyomoviruses*, etc. [3]. This permits to avoid or at least to delay the "*divorse*" of taxonomy based on the use of virus genomes only and of classification systems useful for clinicycts, plant biologists and other specialists (as it was stated in [2]: "*numerous widely used clinical or veterinary virus designations cannot be supported by taxonomic assignments but often better serve clinical and regulatory purposes*").

It seems that such a divorse is partly connected with the difficulty which was mentioned by virologists in [2]: diversity of clinical symptoms is observed for the viruses considered as representing the member of the same species on the basis of the close similarity of their aligned genome sequences. As the example they considered the case of the species *Enterovirus C* which strains have essentially distinct properties. This was considered as a form of "mismatch" that can occur between taxonomy and classification desired by clinicians, veterinarians, agronomists, etc.

The species *Enterovirus C*, in the family *Picornaviridae*, includes a clinically highly diverse range of member viruses, such as poliovirus types 1, 2, and 3 causing poliomyelitis, as well as also other enterovirus types which can cause different diseases of gastrointestinal tract, repiratory and also of nervous system. The assignment of these viruses to the same species is motivated by their high degree of sequence similarity and their ability to recombine [2,6]. Note, that the poliovirus-associated neuroinvasive phenotype ultimately derives from a difference in the receptors used by these viruses, which is caused by only of small difference in the gene encoding the capsid protein VP1.

Here, using the example of *Enterovirus C* viruses, we show that this difficulty disappears when using an approach to constructing a natural viral system using the Neural Replicator Analysis (NRA) [5]. The main advantage of this approach is that it allows the nonlinear preprocessing of original viral genome sequences by converting them into replicator tables and replicator transmitted motifs that exhibit distinct periodic properties [3,4.5].

## 1. Neural Replicator Analysis − brief description

We have already presented in details the approach called Neural Replicator Analysis (NRA) and demonstrated its usefulness in the analysis of viroids [5] and viruses, as well as for constructing a version of natural virus genome table [3,4]. Here we will briefly outline the main concepts of NRA (see [3,5] for further details).

*Neural replicators*

The basic artificial neural network model used in NRA is the self-reproducible neural network (neural replicator) [5, 7, 8]. This model includes the mechanism of synchronously changing threshold of all neurons having binary states $x_i$ (+1 or −1) in the standard Hopfield network [9]. It is suggested that ancestor Hopfield network has arbitrary matrix of interconnections and corresponding set of attractors (stable states) for zero neuron thresholds. This network is placed in a *network ensemble* (e.g., one or two dimensional) consisting of the untrained networks having zero synaptic matrix. The ancestor network can force neighbor network neurons to take values of their neuron states in through one-to-one interconnections in the course of information transmission [7]. The signal of the *start of this transmission* arises when ancestor network puts all the thresholds of their neurons to the very low negative value at once. In this case all states of ancestor network neurons take maximal values (+1). This maximally excited state of ancestor network *opens the channel* of information transmission to neighbor network. Then all thresholds of ancestor network start to grow synchronously taking the same values. At some threshold level the state of some neurons *become unstable* and neural dynamics starts until *equilibrium state* at this threshold will be reached (note, that threshold grow is very slow to permit this process to terminate). This equilibrium (stable) state is transmitted to the neighbor network *forcing it to learn this pattern* with Hebbian rule [9]. Then the growth of thresholds in ancestor network continues and it transmits its quasi stable attractors arisen at different threshold levels to a neighbor network which learns all of them. When the threshold level becomes high enough all neurons become passive (their states take values $x_i = -1$) and this passive network state is interpreted by neighbor network as the signal of the finish of information transmission. After this course the neighbor network learns all quasi stable (stable in given threshold interval) states of ancestor network and becomes a new ancestor network able to transmit information to its untrained neighbor network. So, for example, in linear chain of networks a one-directional wave of learning can be organized. The remarkable phenomena observed in such a system [5, 7, 8] is that after few steps of transmission a special network arises in a chain which transmit further *just those patterns which it*

*learned from its neighbor*. In other words, this network produces its *exact copy*, or is self-reproducible. In effect, identical networks arise and spread through the system. The self-reproducible networks (neural replicators) are absolutely transparent ones − they show as quasi attractors all learned patterns during the cycle of threshold growth.

*Incomplete codes of nucleotide sequence*

The model suggests that neurons take binary values. Though many generalizations of this model permit to avoid this restriction just such code scheme was used for genomic analysis in a previous paper [5]. In this paper *non-traditional representation* of nucleotide sequences was used. Instead of four-letter genetic code *two binary code schemes* to represent these sequences were introduced. The first code (called WS code) combines the Watson-Crick pairs (AT) and (CG) and presents them as a weak (AT) pair encoded by "–1" and a strong (CG) pair encoded by "+1". The second keto-amino (KM) code combines a wobble pair (TG) encoded with "+1" and a less stable (AC) pair encoded with "–1".

*Replicator Tables*

These two incomplete codes were used to construct sets of networks of different sizes $K$ (starting from 3) with the Hebbian interconnections calculated with the use of patterns generated by sliding the nucleotide sequence consisting of $N$ nucleotides with a window having a length $K$. Note, that this can be done easily for viruses having circular genomes considered in [3] and [5]. In the case of linear genomes, e.g. genomes of the genus *Flavivirus* (*Orthoflavivirus*) considered in [4] some approximation can be used. It can be assumed that the NRA also views them as circular, connecting the beginning and end of the chain. This approximation seems to be quite precise, since the maximum length of the sliding window (the maximum number of neurons in the replicator network) is chosen as in [3,4,5] equal to $K=30$. Since the length of viruses of the species *Enterovirus C* considered in this article is about 7 500, this approximation introduces an error of only about 0.4% into the analysis. We also characterize the absence of replicators for $K$ larger than 30 with the abbreviation "NoR" (No Replicators), which means that they really do not exist for networks of all considered sizes of $K \leq 30$. The N resulting patterns are then used to form the Hebbian connectivity matrices of the two parent fully connected Hopfield networks (for WS- and KM-encoded patterns, respectively). Next, using the method described above, self-reproducing replicators are obtained. For simplicity and to avoid the ambiguity (the appearance of different sets of replicators) asynchronous but ordered dynamics for updates of the states of neurons in the Hopfield network is suggested. The results of studies are presented in Replicator Table (RT )[5] which presents the presence or absence replicators for both code schemes (WS and KM) and different network size, $K$. Replicator sets have been shown to differ significantly between two partial representations of viroid nucleotide sequences (derived using the WS and KM codes) [5], as well as for different viral genomes [3,4]. The simplest difference is the fact that for a sliding window of the same size the source parent network can generate a nontrivial replicator with a non-empty set of the patterns for transmission, or non-replicating network with empty set of patterns for transmission. This last network cannot generate descendants or, in other words, cannot breed.

*Fuzzy motifs*

The remarkable phenomenon is connected to the replicator transmitted patterns - fuzzy motifs. It was shown that patterns transmitted by replicators contain additional information and often have interesting symmetries and periodicities [3,4,5]. This fact was used for building natural virus genome system in the form of table [4]. This demonstrated the potential usefulness of the NRA for constructing a binomial classification of virus genomes based only on knowledge of their complete genomic sequences, without involving other data on phenotype, functions, encoded proteins, etc., and also without the need to align genomic sequences. Comparison of genomic sequences plays an important role in the taxonomy of viruses to distinguish between types, species, genera and families, so NRA applied to virus genomes can, in principle, provide some additional information for virus classification. We have demonstrated in [3] that the periodicity of replicator patterns (fuzzy motifs) can be used to organize most viral genomes into a rectangular table to obtain their binomial classification.

The key factor is the possibility of independent analysis of two binary sequences representing WS-encoded and KM-encoded genomes. This approach makes it possible to combine the genomes of viruses that are far from each other in terms of the similarity of aligned nucleotide sequences (as in the case of representatives of the $\alpha$- and $\nu$-human papillomaviruses, and even in the case of viruses belonging to different kingdoms, such as the crow polyomavirus and allamanda leaf mottle distortion virus [3]) or, on the other hand, separate them when they have similar aligned genomes), as in the case of $\nu$ human papillomavirus and porcupine papillomavirus $\sigma$.

We have also demonstrated [3] that NRA may, to some extent, reflect such general characteristics of the viral phenotype as:

- virus hosts and their possible interconnections (e.g. for the genus *Badnavirus* and for the family *Polyomaviridae* [3]);
- the form of disease (e.g., mosaic disease or yellow mottle disease caused by members of genus *Badnavirus*);
- the virus vector (e.g. tick-borne or mosquito-borne viruses from the genus *Flavivirus* [4]);
- oncogenicity of the virus (tendency to the absence of neural replicators for both genome coding schemes);

- the form of the epithelium affected by the virus (skin or mucous) (for the *human papillomavirus* [3]);
- morphology (members of the genus *Badnavirus* with bacillary geometry are presumably located in the same column of virus genome table [3]).

## 2. Neural Replicator Analysis of *Enterovirus C* species

Now, we use NRA to demonstrate that it can reveal simple fine structure of the species *Enterovirus C*. All viruses of this species do not have replicators for KM-encoded genomes, similar to *Human papilloma viruses* and the vast majority of *Polyomaviruses* [3]. Half of these viruses, when their genomes are WS-encoded, have replicators containing only one motif for some network sizes $K$. These motifs are periodic with period values T=2, 3 and 4 (see Table 1, where these viruses are highlighted in color). Therefore, the corresponding viruses can be placed in three cells of the viral binomial table [3] and, remarkably, they cause diseases in various human systems: the nervous system (green), the gastrointestinal tract (yellow) and the respiratory system (blue). This may make the NRA and the corresponding genome table of natural viruses useful tools for clinicians. Other viruses (shown in gray in Table 1) have a generally complex set of replicator motifs, which are characterized by a mixture of periodicities of different replicator patterns (motifs) and a mixture of periodicities within a single pattern. In addition to T=2,3 and 4, there are also motifs with a periodicity of T=5 and also T=7. When the first part of viruses which we can call basis ones can be easily placed in the virus genome table these last viruses with mixed periodicities may be considered as combinations of the basis ones. They can be used as indication of the existence of other basis viruses with longer period values, e.g., with the periods to T=5,7.

Let us now consider basic viruses with motif periods T = 2, 3 and 4, shown in Fig. 1. Eleven *Enterovirus C* viruses have replicators containing only one motif for WS-encoded genomes. The viruses for which T=2 are three polioviruses and two coxsackieviruses, which cause aseptic meningitis. Thus, cellular viruses (2, NoR) cause diseases of the nervous system (other polioviruses have mixed periods, but they also have 2-periodicity). The strains for which T = 3 are two coxsackieviruses and two enteroviruses that cause gastroenteritis. Thus, cellular viruses (3, NoR) cause diseases of the gastrointestinal tract. Finally, viruses for which T=4 belong to the cell (4, NoR) are two enteroviruses that cause respiratory diseases, including pneumonia. Examples of replicator tables of three viruses belonging to these three different cells and the periodic motifs of replicators are shown in Fig. 2.

Thus, Neural Replicator Analysis separates viruses of *Enterovirus C* species into at least on three major subspecies, which likely correspond to the types of *tissues* damaged by the viruses. This is reminiscent of NRA's success in separating *Human papillomaviruses* species belonging to the *genus Alpha*, for which it has been shown that, unlike other species that are particularly *oncogenic* and cause *mucosual* lesions, species associated with warts which are *cutaneous* lesions have 2-periodic motifs and belong to cell (2, NoR) [3].

The locations of the already known viruses, as well as potential ones, are shown in the combined table of viruses presented in Fig. 3 (compare with the earlier version presented in [4]).

Table 1. Viruses of the *Enterovirus C* species studied in this article. Replicators with single motifs that have clearly defined single periods are colored. The remaining replicators are shown in gray. The green, yellow and blue colors correspond to viruses that cause diseases of the nervous system, gastrointestinal tract and respiratory system, respectively

| Type | Abbrev | Isolate | Accession | Replicator maximun size | Motis periods |
|---|---|---|---|---|---|
| **Poliovirus 1** | **PV1** | **Mahoney** | **V01149** | **26** | **2** |
| **Poliovirus 1** | **PV1** | **Sabin (L5c-2ab)** | **V01150** | **23** | **2** |
| **Poliovirus 2** | **PV2** | **Lancing(Michigan/37)** | **M12197** | **26** | **2** |
| Poliovirus 2 | PV2 | Sabin (P712-Ch-2ab) | X00595 | 19 | 2, 5 |
| Poliovirus 3 | PV3 | Leon (California/37) | K01392 | 30 | 2, 3 |
| Poliovirus 3 | PV3 | Sabin (Leon 12a-1-b) | X00925 | 30 | 2, 3 |
| **Coxsackievirus A1** | **CVA-1** | **T.T. (Tompkins)(Coxsakie/NY/47)** | **AF499635** | **7** | **3** |
| **Coxsackievirus A11** | **CVA-11** | **Belgium 1 (Belgium/51)** | **AF499636** | **20** | **3** |
| Coxsackievirus A13 | CVA-13 | Flores (Mexico/52) | AF499637 | 18 | 2,3 |
| Coxsackievirus A18 | CVA-18 | G-13 (South Africa/50) | AF499640 | 13 | 2,3 |
| Coxsackievirus A17 | CVA-17 | G-12 (South Africa/51) | AF499639 | 22 | 2,3 |
| **Coxsackievirus A19** | **CVA-19** | **NIH-8663 (Dohi)(Japan/52)** | **AF499641** | **6** | **2** |
| Coxsackievirus A21 | CVA-21 | Kuykendall (California/52) | AF546702 | 20 | 2,3,5,7 |
| **Coxsackievirus A22** | **CVA-22** | **Chulman (NewYork/55)** | **AF499643** | **10** | **2** |
| Coxsackievirus A24 | CVA-24 | Joseph | EF026081 | 22 | 2,3 |
| Enterovirus C96 | EV-C96 | BAN00-10488 | EF015886 | 13 | 2,3 |
| Enterovirus C99 | EV-C99 | USA-GA84-10636 | EF555644 | 16 | 3,5 |
| Enterovirus C102 | EV-C102 | BAN99-10424 | EF555645 | 13 | 2,3 |
| Enterovirus C105 | EV-C105 | PER153 (Peru/2010) | JX393302 | 30 | 2,3,5,7 |
| **Enterovirus C109** | **EV-C109** | **NICA08-4327** | **GQ865517** | **23** | **4** |
| **Enterovirus C113** | **EV-C113** | **BBD-48 (Bangladesh/2009)** | **KC344833** | **11** | **3** |
| **Enterovirus C116** | **EV-C116** | **126/Russia/2010** | **JX514942** | **11** | **3** |
| **Enterovirus C117** | **EV-C117** | **LIT22/Vilnus (Lithuania/2011)** | **JX262382** | **30** | **4** |
| Enterovirus C118 | EV-C118 | ISR10 (Israel/2011) | JX961708 | 29 | 2,3,5 |

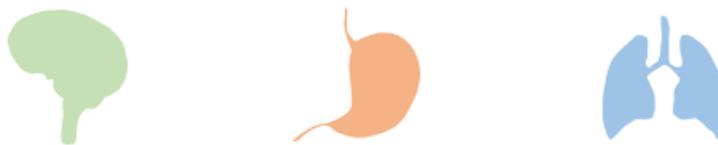

**Fig. 1.** Eleven *Enterovirus C* viruses have replicators containing only one motif for WS-encoded genomes. These motifs are periodic with period values T=2,3 and 4. Viruses shown in each of these cells cause diseases of the nervous system, gastrointestinal tract and respiratory system, respectively.

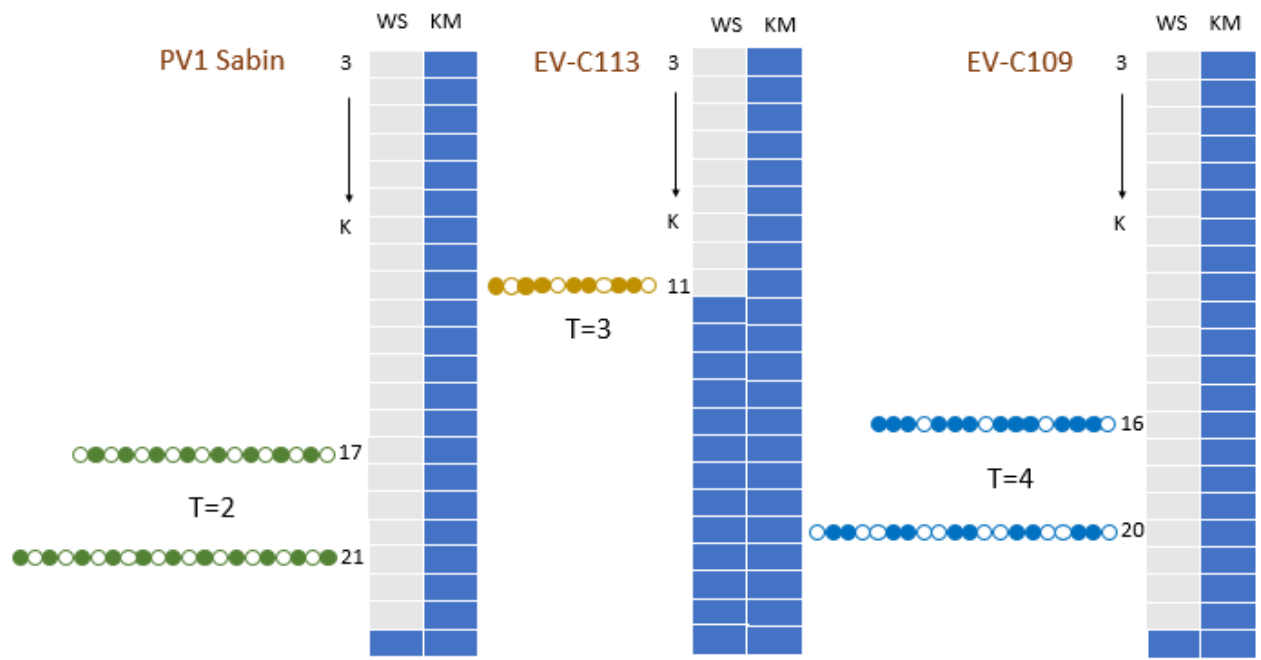

**Fig. 2.** Replicator Tables of three strains of *Enterovirus C*, located in 3 different cells of the virus genome table − (2, NoR), (3, NoR), (4, NoR) - and having single periodic replicator motifs: (T=2, for K=17, 21, Sabin strain PV1); (T=3, for K=11, EV-C113); (T=4, for K=16, 20, EV-C109). Motifs are represented by a chain of circles where a white circle denotes a neuron state of +1, while colores circle denotes neuron state of −1. Green, yellow and blue colors in the motifs correspond to diseases of nervous system, gastrointestinal tract and respiratory system, correspondingly. In the Replicator Table, gray rectangle indicates the existence of a replicator for a given neural network size *K*, while blue rectangle indicates the absence of a replicator (see also [3,4,5] for more details).

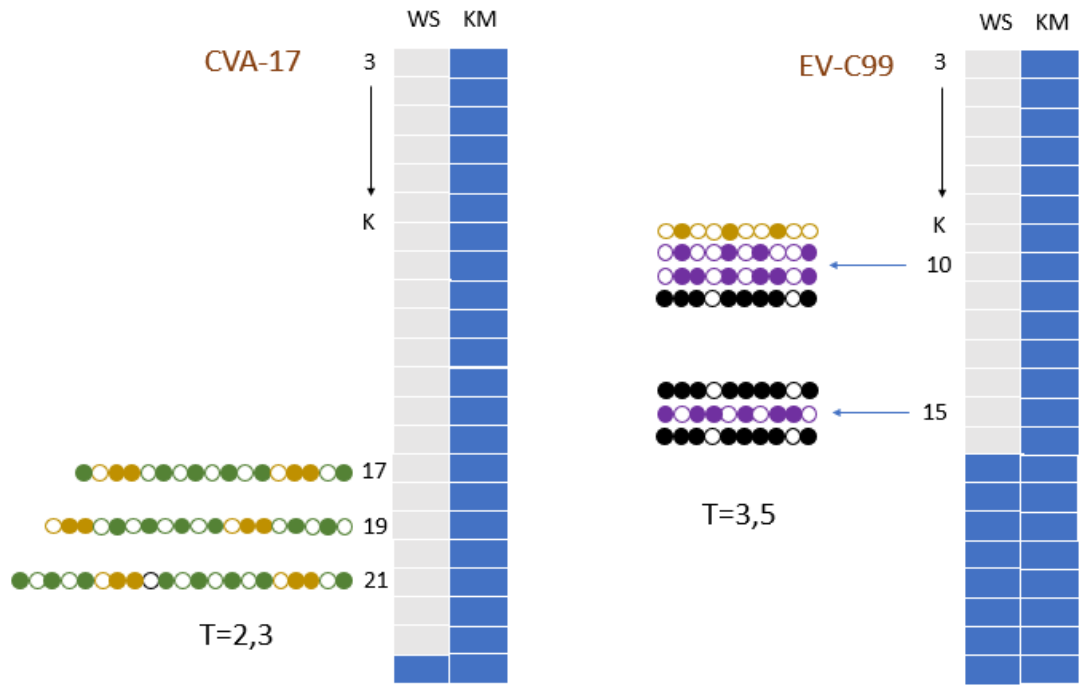

**Fig. 3.** Replicator tables of *Enterovirus C* strains CVA-17 (left) and EV-C99 (right). Single motif CVA-17 replicators with K=17, 19, 21 exhibit an intermittency of 2- and 3-periodicitity, which can also be considered as 11-periodicity. EV-C99 replicator motif sets with K=10, 15 contain motifs with 3- and 5-periodicity.

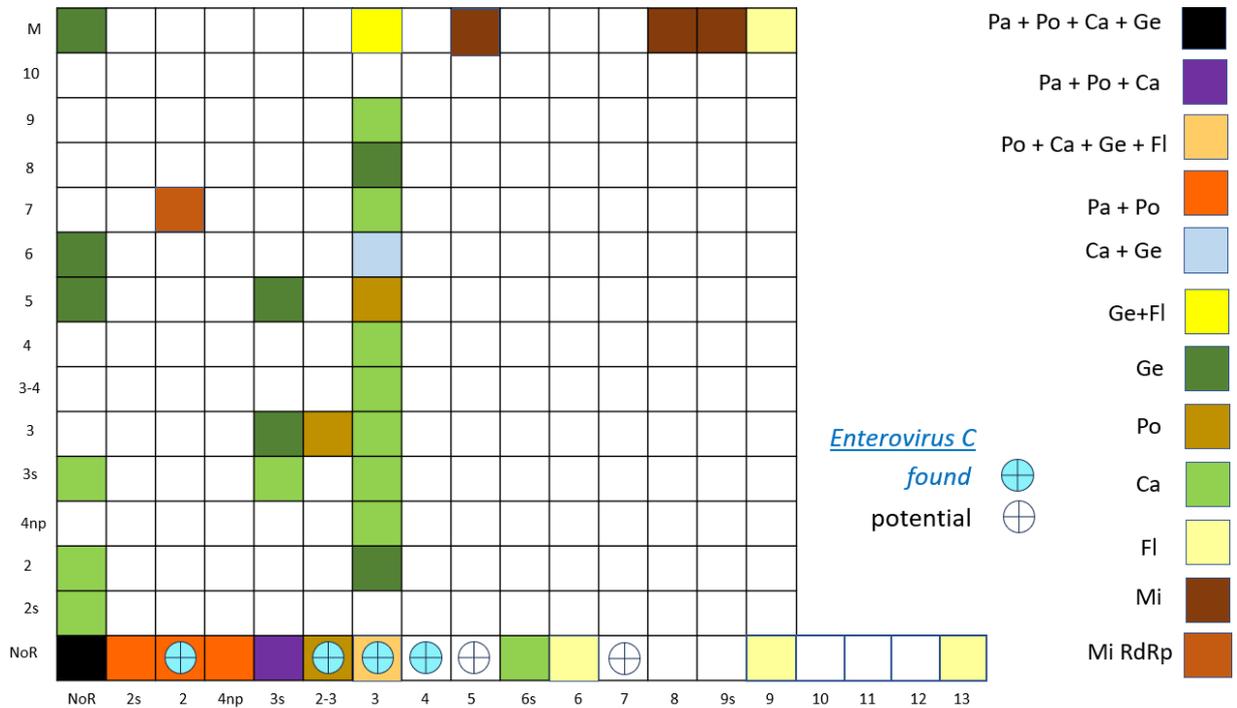

**Fig. 4.** Joint table of viral genomes showing cells populated according to NRA data obtained for the genus *Flavivirus* (Fl) [4] and also presented in [3] families *Papillomaviridae* (human) – Pa, *Polyomaviridae* – Po, *Caulimoviridae* – (Ca), *Geminiviridae* – Ge and also *Mitoviridae* – Mi (Gigaspora margarita mitovirus 1, NC_040702.1 – (5, M), Cronartium ribicola mitovirus 5, NC_030399.1 –(8, M); Fusarium poae mitovirus 4, NC_030864.1 – (9s, M) ; Rhizoctonia mitovirus 1 RdRp, NC_040563.1 – (2, 7) ). Some cells are filled with genomes of viruses belonging to different families. *Enterovirus C* viruses with a single period of transmitted patterns (fuzzy motifs) are located in already colonized cells (2, NoR), (3, NoR) [3,4], as well as in new cells (4, NoR) and potentially in (5, NoR) and (7, NoR). Another cell occupied by *Enterovirus C* viruses with mixed motifs (2-3, NoR) is also shown.

## 3. Discussion

The first principle of virus taxonomy formilated in [2] requires that virus taxa should be *monophyletic*. In the ICTV taxonomy, classification is *evolutionarily* based and *hierarchical*. But in general, biological classification can be constructed without reference to evolution, and does not necessarily have to be hierarchical. The use of a *tabular form* to represent biological objects instead of hierarchical trees was discussed in the studies of Alexander A. Lubischew [10]. Lubischew argued that a *natural system* in biology may have a form that does not reflect the evolution of species, but be similar to Mendeleev's periodic table of chemical elements.

Also, it does not have to be obligatory monophyletic or, in general, monothetic This is also consistent with van Regenmortel definition of the virus species. In 1991, ICTV adopted his formulation, which defines a virus species as a *polythetic class* of viruses that constitute a replicating lineage and occupy a particular ecological niche" [11, 12]. A polythetic class consists of members that have a number of common properties, but not all have one common property [11].

According to Lubischew, the properties of biological objects should follow from their position in the table depicting their natural system. We proposed some version of this natural tabular system for viral genomes in [3]. The genome position in this table has two coordinates, which are determined by the periodicity of the motifs transmitted by neural replicators, constructed using two incomplete representations of nucleotide sequence based of the WS and KM codes [3]. Therefore, this natural system is also binomial. For example, cell (2,3) defines viral genomes whose motifs have 2-periodicity for WS-encoded nucleotide sequence and 3-periodicity for KM-encoded nucleotide sequence. It has been shown [3] that this position in the table actually determines in many cases the different phenotypic properties of viruses. Note, that neural replicators are built on the basis of energy-minimizing Hopfield neural networks [9] with different neural thresholds, and the motifs of the replicators can be interpreted as prototypes, the meaning of which corresponds to the polythetic nature of pattern classification by the Hopfield network [13].

It can be concluded that analysis of neural replicators can divide *Enterovirus C* species into subspecies that correspond to different types of diseases caused by *Enterovirus C* viruses. This allows the tabular form of the natural viral system to be used to reconcile information on only viral genomes (as occurs in metagenomic studies) with properties of the virus that are useful for clinicians.

**Acknowledgments** I am grateful to Maria Mishina, who prepared the drawings for this article